\documentclass[fleqn,usenatbib]{mnras}

\usepackage{newtxtext,newtxmath}
\usepackage{float}

\usepackage[T1]{fontenc}

\DeclareRobustCommand{\VAN}[3]{#2}
\let\VANthebibliography\thebibliography
\def\thebibliography{\DeclareRobustCommand{\VAN}[3]{##3}\VANthebibliography}

\usepackage{pstricks, pst-plot}	
\usepackage{graphicx} 
\usepackage{wrapfig}  
\usepackage[figuresright]{rotating}
\usepackage{subcaption}
\usepackage{float}

\usepackage{soul}

\usepackage{graphicx}	
\usepackage{amsmath}	

\title[Equation of state for hot hyperonic neutron star matter]{ Equation of state for hot hyperonic neutron star matter}

\author[H. Kochankovski et al.]{
Hristijan Kochankovski,$^{1,2}$\thanks{E-mail: hriskoch@fqa.ub}
Angels Ramos,$^{1}$\thanks{E-mail: ramos@fqa.ub.edu}
Laura Tolos$^{3,4,5}$\thanks{E-mail: tolos@ice.csic.es}
\\
$^{1}$Departament de F\'{\i}sica Qu\`antica i Astrof\'{\i}sica and Institut de Ci\`encies del Cosmos, Universitat de Barcelona, Mart\'i i Franqu\`es 1, 08028, Barcelona, Spain\\
$^{2}$Faculty of Natural Sciences and Mathematics-Skopje, Ss. Cyril and Methodius University in Skopje, Arhimedova, 1000 Skopje, North Macedonia \\
$^{3}$Institute of Space Sciences (ICE, CSIC), Campus UAB,  Carrer de Can Magrans, 08193 Barcelona, Spain\\
$^{4}$Institut d'Estudis Espacials de Catalunya (IEEC), 08034 Barcelona, Spain\\
$^{5}$Frankfurt Institute for Advanced Studies, Ruth-Moufang-Str. 1, 60438 Frankfurt am Main, Germany
}

\date{Accepted XXX. Received YYY; in original form ZZZ}

\pubyear{2022}

\begin{document}
\label{firstpage}
\pagerange{\pageref{firstpage}--\pageref{lastpage}}
\maketitle

\begin{abstract}
The FSU2H equation-of-state model, originally developed to describe cold neutron star matter with hyperonic cores, is extended to finite temperature. Results are presented for a wide range of temperatures and lepton fractions, which cover the conditions met in protoneutron star matter, neutron star mergers and supernova explosions. It is found that the temperature effects on the thermodynamical observables and the composition of the neutron star core are stronger when the hyperonic degrees of freedom are considered. An evaluation of the temperature and density dependence of the thermal index leads to the observation that the so-called $\Gamma$ law, widely used in neutron star merger simulations, is not appropriate to reproduce the true thermal effects, specially when hyperons start to be abundant in the neutron star core. To make finite temperature equations of state easily accessible, simple parameterizations of the thermal index for nucleonic and hyperonic $\beta$-stable neutrino-free matter are provided. 
\end{abstract}

\begin{keywords}

stars:neutron -- dense matter -- equation of state
\end{keywords}



\section{Introduction}

\label{intro}


Neutron stars are one of the most compact objects in the universe. Due to their extreme properties, that is, radii around $11-15$ km (see latest results from the NICER collaboration \cite{Riley:2019yda,Miller:2019cac,Riley:2021pdl,Miller:2021qha}) and masses that can exceed $2M_{\odot}$ \cite{Demorest2010ShapiroStar,Antoniadis:2013pzd,Fonseca2016,NANOGrav:2019jur}, they are a natural laboratory for studying matter under extreme conditions. In particular, the core of the neutron star is the most intriguing part as very little is known about its composition, whether only nucleonic degrees of freedom are present or more exotic components can appear.

The description of the cold neutron star core is given by one-parameter equation of state (EoS) that relates the pressure to (energy) density. On the contrary, when one considers the evolution of a young neutron star \cite{Pons1999EvolutionStars,Pascal:2022qeg}, the collapse of supernovae  \cite{mezzapaca:corrected,Janka:2006fh,2017RSPTA.37560271C,OConnor2018ExploringSupernovae,Burrows:2019zce} or the merger of a binary system of neutron stars (see, for example, \cite{Rosswog:2015nja,Baiotti:2016qnr}), a finite temperature treatment is necessary and, hence, the EoS depends on three parameters, that is, temperature $T$, lepton fraction $Y_l$ and baryon density $\rho_B$. In order to account for the different conditions in the aforementioned astrophysical events, the core EoS needs to cover a wide range of these parameters: $T = 0-100$ MeV, $\rho_B = \rho_0/2 -8\rho_0$\footnote{$\rho_0$ is the nuclear saturation density of a value of $\rho_0 \sim 0.16 \ {\rm fm^{-3}}$}, $Y_l = 0 - 0.4$ \cite{Oertel:2016bki}.

Under these conditions of temperature, density and leptonic fraction, the appearance
of non-nucleonic degrees of freedom in the hot core, such as hyperons, meson condensates or quarks, is energetically probable. This is due to the high value of density at the center of the star and the rapid increase of the nucleon chemical potential with density, that makes nucleons energetically less favoured in comparison with more exotic species.

For almost two decades only two models of the so-called general purpose EoS for astrophysical events existed, that is, the one by Lattimer and Swesty \cite{Lattimer:1991nc} and that of Shen et al. \cite{Shen:1998by,Shen:1998gq}, which only considered nucleons.  Indeed, EoSs accounting for finite temperature corrections were built using the $\Gamma$ law, that consists in implementing thermal contributions from the ideal gas to the cold EoS. The situation clearly ameliorated in the last decade, with almost one hundred general purpose EoSs, and about one third considering exotic degrees of freedom (see reviews of \cite{Oertel:2016bki, Burgio:2021vgk,Typel:2022lcx}). Though some progress has been done (see \cite{Raduta:2021coc,Raduta:2022elz} and references therein), the number of models is still far from desirable, due to the large parameter space associated with the uncertainties in the effective interactions at high densities and isospin asymmetry.

With the present work we aim at improving the present situation by reexamining our FSU2H model for the nucleonic and hyperonic core of neutron stars, that satisfies the $2M_{\odot}$ as well as produces stellar radii below 13 km, while
fulfilling the saturation properties of nuclear matter and finite nuclei together with the constraints
on the high-density nuclear pressure coming from heavy-ion collisions \cite{Tolos:2016hhl,Tolos:2017lgv}.

We first focus on improving some of its features at zero temperature. This will be done by, on the one hand, modifying the coupling of the $\sigma$ field to the $\Xi$ baryon so as to reproduce the latest value of the $\Xi$ nuclear potential \cite{Friedman:2021rhu}, and, on the other hand, by incorporating the $\sigma^*$ field in order to have a better description of the $\Lambda \Lambda$ bond in $\Lambda$ matter \cite{Takahashi:2001nm,Ahn:2013poa}. We will refer to this model as FSU2H$^*$.

The second goal of the present work is to provide a better understanding of finite-temperature behavior of EoS models with exotic components. This will be done by extending the new FSU2H$^*$ model at finite temperature. In particular, we are interested in determining the validity of the $\Gamma$ law when exotic degrees of freedom are present in the core of hot neutron stars. The final objective is to analyze the phenomena of supernovae and neutron star binary mergers with a solid understanding of the interior of neutron stars. 

The paper is organized as follows. First, in Section \ref{sec:1} we present the theoretical framework, while introducing the new FSU2H$^*$ model. The formalism at finite temperature is discussed afterwards. Then, in Section \ref{sec:2} we show the composition and the EoS for different values of $T$ and $Y_l$, while paying a special attention to the thermal index. We specifically address the differences arising when hyperons are taken into account, while discussing the importance of an exact treatment of the finite temperature calculations beyond the $\Gamma$ law. In order to make the EoS usable for numerical simulations, we also provide a simple functional dependence of the thermal index for the neutrino free case. Finally, in Section \ref{sec:3} a brief summary is provided.

\section{Theoretical framework}
\label{sec:1}
We consider matter made of baryons and leptons at a given temperature $T$, baryon  density $\rho_B$ and fixed lepton fraction $Y_l$. Within the relativistic field theory, the interaction between the baryons is modeled through exchange of different mesons \cite{Walecka:1974qa, Boguta1977RelativisticSurface,Serot1997RecentHadrodynamics,Glendenning:2000,Chen:2014sca}. The Lagrangian density $\cal{L}$ of the system may be split into baryonic contributions ${\cal L}_b$ 
($b$ =  $n, p,\Lambda$, $\Sigma$, $\Xi$), a mesonic term ${\cal L}_m$, which includes the contributions from the $\sigma$ , $\omega$, $\rho$, $\phi$ and $\sigma^{*}$ mesons, and leptonic contributions ${\cal L}_l$ ($l$ =  $e,\mu$ and the corresponding neutrinos) as
\begin{eqnarray}
{\cal L} &=& \sum_{b} {\cal L}_b + {\cal L}_m +\sum_{l} {\cal L}_l ,\nonumber \\
{\cal L}_b &=& \bar{\Psi}_b(i\gamma_{\mu}\partial^{\mu} -q_b{\gamma}_{\mu} A^{\mu} - m_b \nonumber \\
&+& g_{\sigma b}\sigma + g_{\sigma^{*}b}  \sigma^{*} - g_{\omega b}\gamma_{\mu} \omega^{\mu}\nonumber \\
&-& g_{\phi b}\gamma_{\mu} \phi^{\mu} - g_{\rho,b}\gamma_{\mu}\vec{I}_{b}\cdot \vec{\rho\,}^{\mu})\Psi_b, \nonumber \\
{\cal L}_m &=& \frac{1}{2}\partial_{\mu}\sigma \partial^{\mu}\sigma - \frac{1}{2}m^2_{\sigma}\sigma^2 - \frac{\kappa}{3!}(g_{\sigma b}\sigma)^3 - \frac{\lambda}{4!}(g_{\sigma b})^4 
 \nonumber \\ 
&+& \frac{1}{2}\partial_{\mu}\sigma^{*} \partial^{\mu}\sigma^{*}  -\frac{1}{2}m^2_{\sigma^{*}}{\sigma^{*}}^2 \nonumber \\
&-&\frac{1}{4}\Omega^{\mu \nu}\Omega_{\mu \nu}  
+\frac{1}{2}m^2_{\omega} \omega_{\mu} {\omega}^{\mu} +  \frac{\zeta}{4!} g_{\omega b}^4 (\omega_{\mu}\omega^{\mu})^2 \nonumber \\
&-&\frac{1}{4}\vec{R}^{\mu \nu}\cdot \vec{R}_{\mu \nu} + \frac{1}{2}m^2_{\rho}\vec{\rho}_{\mu}\cdot\vec{\rho\,}^{\mu}+ 
\Lambda_{\omega}g^2_{\rho b}\vec{\rho_{\mu}}\cdot\vec{\rho\,}^{\mu} g^2_{\omega b} \omega_{\mu} \omega^{\mu} \nonumber \\
&-& \frac{1}{4}P^{\mu \nu}P_{\mu \nu}
+\frac{1}{2}m^2_{\phi}\phi_{\mu}\phi^{\mu}-\frac{1}{4}F^{\mu \nu}F_{\mu \nu} , \nonumber \\
{\cal L}_l &=& \bar{\Psi}_l\left(i\gamma_{\mu}\partial^{\mu}-q_l\gamma_{\mu}A^{\mu} - m_l \right)\Psi_l,  
\label{eq:lagrangian}
\end{eqnarray}
 with $m_i$ being the mass of $i$-th particle, $\Psi_b$ and $\Psi_l$ the baryon and lepton Dirac fields, respectively, while $\Omega_{\mu \nu} = \partial_{\mu} \omega_{\nu} -\partial_{\nu} \omega_{\mu} $, $\vec{R}_{\mu \nu} = \partial_{\mu} \vec{\rho_{\nu}} - \partial_{\nu} \vec{\rho_{\mu}} $, $P_{\mu \nu} = \partial_{\mu} \phi_{\nu} -\partial_{\nu} \phi_{\mu} $ and $F_{\mu \nu} = \partial_{\mu} A_{\nu} -\partial_{\nu} A_{\mu}$ are the mesonic and electromagnetic strength tensors. Lastly, with $\vec{I}_b$ we represent the isospin operator, $\gamma^{\mu}$ are the Dirac matrices and $g_{mb}$ labels the couplings of the different baryons to the mesons. We note that the $\sigma, \omega$ and $\rho$ mesons mediate the interaction between any type of baryons in the octet, while the $\sigma^{*}$ and $\phi$ mediate the interaction only between those having non-zero  strangeness. The coupling constants encode the complicated nuclear many-body dynamics, and they are taken so that the  saturation properties of nuclear matter and finite nuclei, together with the constraints on the high-density nuclear pressure coming from heavy-ion collisions, as well as the 2 $M_{\odot}$ neutron star observations and radii smaller than 13 km, are well reproduced. For an extensive discussion of the role played by the different coupling constants in our model see \cite{Tolos:2016hhl,Tolos:2017lgv}. As compared to \cite{Tolos:2016hhl,Tolos:2017lgv}, we have introduced the exchange with the $\sigma^*$ so as to improve our description of hyperonic potentials, as we will discuss in the following.
 
In order to obtain the thermodynamic properties and composition of matter at finite temperature under $\beta$-equilibrium conditions, one starts by determining
the Dirac equations for the different baryons and leptons from the Lagrangian density of Eq.~(\ref{eq:lagrangian}):
\begin{eqnarray}
    &&(i\gamma_{\mu}\partial^{\mu} - q_{q}\,\gamma_{\mu} \,A^{\mu} - m_b^{*} - g_{\omega b}\gamma_0 \omega^0 - g_{\phi b}\gamma_{0}\phi^0 \nonumber \\
    &&\phantom{(i\gamma_{\mu}\partial^{\mu} - q_{q}\,\gamma_{\mu} \,A^{\mu} - m_b^{*} - g_{\omega b}\gamma_0}- g_{\rho b} I_{3b}\gamma_0\rho_3^0)\Psi_b = 0, \nonumber \\
    &&\left(i\gamma_{\mu}\,\partial^{\mu}-q_{l}\,\gamma_{\mu} \,A^{\mu}-m_{l} \right) \psi_{l}=0 ,
    \label{eq:baryons-leptons}
\end{eqnarray}
with the effective mass of the baryons given by
\begin{equation}
    m_b^{*} = m_b - g_{\sigma b} \sigma- g_{\sigma^{*} b} \sigma^{*},
    \label{eq:meff}
\end{equation}
and $I_{3b}$ being the third component of the isospin of a given baryon, with the convention that for protons $I_{3p} = +1/2$. Note that only the time-like component of the vector fields and the third component of isospin have been written in Eq.~(\ref{eq:baryons-leptons}) due to the assumption of rotational invariance and charge conservation.

The field equations of motion for the mesons are obtained from the Euler-Lagrange equations (see, for example, \cite{Serot:1984ey}). Within the relativistic mean-field approximation, the meson field operators are replaced by their expectation values. By denoting  $\bar \sigma = <\sigma>$, $\bar \rho = <\rho_3^0>$, $\bar \omega = <\omega^0>$,  $\bar \phi = <\phi^0>$ and $\bar \sigma^{*} = <\sigma^{*}>$ we obtain
\begin{eqnarray}
&&m_{\sigma}^2\bar \sigma + \frac{\kappa}{2}g_{\sigma b}^3 \bar \sigma^2 + \frac{\lambda}{3!}g_{\sigma b}^4 \bar \sigma^3 = \sum_{b} g_{\sigma b} \rho_b^s , \nonumber \\
&&m_{\sigma^{*}}^2 \bar \sigma^{*} = \sum_{b} g_{\sigma^* b} \rho_{b}^s, \nonumber \\
&&m_{\omega}^2 \bar \omega + \frac{\zeta}{3!}g_{\omega b}^4 \bar \omega^3 + 
2 \Lambda_{\omega} g^2_{\rho b}g^2_{\omega b} \bar \omega \bar \rho^2 = \sum_{b}  g_{\omega b} \rho_b , \nonumber \\
&&m_{\rho}^2 \bar \rho +  
2 \Lambda_{\omega} g^2_{\rho b}g^2_{\omega b} \bar \omega^{2} \bar \rho = \sum_{b} \ g_{\rho b}I_{3b} \rho_b , \nonumber \\
&&m_{\phi}^2 \bar \phi = \sum_{b} g_{\phi b} \rho_b ,
\label{eq:mesons}
\end{eqnarray}
where the scalar $\rho_b^s$ and vector $\rho_b$ 
densities at finite temperature are given by
\begin{eqnarray}
\label{eq:bary_dens}
\rho_b&=& <\bar{\Psi}_b \gamma^0 \Psi_b> = \frac{\gamma_b}{2 \pi^2}\int_0^{\infty}\!\! dk \, k^2 \, (f_{b}(k,T) - f_{\bar{b}}(k,T))  , \nonumber \\
\rho_b^s&=& <\bar{\Psi}_b\Psi_b> = \frac{\gamma_b}{2 \pi^2}\int_0^{\infty} \!\! dk \, k^2 
\, \frac{m^*_b}{\sqrt{k^2+m_b^{*2}}}\left(f_{b}(k,T)+f_{\bar{b}}(k,T)\right),
\end{eqnarray}
with $\gamma_b=2$ accounting for the degeneracy of the spin degree of freedom of baryons and 
\begin{equation}
f_{b/\bar{b}}(k,T) =\left[1+\text{exp}\left(\frac{\sqrt{k^2+m_b^{*2}} \mp \mu_b^{*}}{T}\right)\right]^{-1},
\label{eq:distribution}
\end{equation}
being the Fermi-Dirac distribution for the baryon ($b$) and antibaryon ($\bar b$) with an effective mass $m_b^{*}$, and the corresponding effective chemical potential given by
\begin{equation}
\mu_b^{*} = \mu_b - g_{b\omega}\bar \omega  - g_{b\rho} I_{3b} \bar \rho- g_{b\phi}\bar \phi.
\label{eq:mueff}
\end{equation}
 We have followed the standard derivation for finite temperature within RMF models as described, for example, in Ref.~\cite{Shen:1998gq}.

The equations for the baryon, lepton and meson fields have to be solved simultaneously  with the condition of $\beta$-stability, which is achieved by imposing the following relations among the chemical potentials of the different species
\begin{eqnarray}
\label{eq:beta_chemical_potentials}
&&\mu_{b^0} = \mu_n , \nonumber \\
&&\mu_{b^{-}} = 2\mu_n - \mu_p , \nonumber \\
&&\mu_{b^{+}} = \mu_p ,\nonumber \\
&&\mu_n - \mu_p = \mu_e - \mu_{\nu_e},\nonumber \\
&&\mu_e = \mu_{\mu} + \mu_{\nu_e} +\mu_{\bar{{\nu}}_{\mu}} ,
\end{eqnarray}
where $b^0$, $b^{-}$, $b^{+}$ indicate neutral, negatively charged and positively charged baryons, respectively. Moreover, $\beta$-stable matter is charge neutral, requiring
\begin{equation}
0=\sum_{b,l} q_i \rho_i ,
\end{equation}
where $q_i$ denotes the charge of the particle $i$.
Finally, the baryon and lepton densities are also conserved
\begin{eqnarray}
&&\rho_B = \sum_{b} \rho_b, \nonumber \\\
&&Y_l \cdot \rho_B = \rho_{l}+ \rho_{\nu_l} ,
\end{eqnarray}
where $\rho_B$ is the total baryon density, $Y_l$ is the lepton fraction for a given flavour, and $\rho_{l(\nu_l)}$ the lepton (leptonic neutrino) density:
\begin{equation}
\rho_{l (\nu_l)}=\frac{\gamma_{l(\nu_l)}}{2 \pi^2}\int_0^{\infty} dk \, k^2 \, {(f_{l (\nu_l)}(k,T) - f_{\bar{l} (\bar{\nu}_l)}(k,T) )},
\end{equation}
with $f_{l(\nu_l)} (k,T)$ and {$f_{\bar l(\bar \nu_l)} (k,T)$} the Fermi-Dirac distribution for leptons and neutrinos and their antiparticles, respectively.

Note that in an extreme neutrino-free case $\mu_{\nu_l}={\mu}_{\bar{\nu}_l}=0$ and the lepton number is no longer a conserved quantity. 

\begin{table*}
\caption{Parameters of the model FSU2H*. The mass of the nucleon is equal to $m_N = 939$ MeV.}
\label{tab_2_1}       
\begin{tabular}{cccccccccccccc}
\hline\noalign{\smallskip}
$m_{\sigma}$  & $m_{\omega}$ & $m_{\rho}$&$m_{\sigma^{*}}$ & $m_{\phi}$ &$g_{\sigma N}^2$  & $g_{\omega N}^2$ & $g_{\rho N}^2$ & $\kappa$ & $\lambda$ & $\zeta$ & $\Lambda_{\omega}$ \\
(MeV) &  (MeV)&  (MeV)&  (MeV)& (MeV) & & & & (MeV) & & &\\
\noalign{\smallskip}\hline\noalign{\smallskip}
497.479 & 782.500 & 763.000 & 980.000  & 1020.000 & 102.72 & 169.53 & 197.27 & 4.00014 & -0.0133 & 0.008 & 0.045 \\
\noalign{\smallskip}\hline
\end{tabular}
\label{table:nuclearparam}
\end{table*}

The different thermodynamic variables can be then straightforwardly obtained from the stress-energy momentum tensor 
\begin{equation}
T_{\mu \nu} =  \frac{\partial {\cal L}}{\partial (\partial_{\mu}\Phi_{\alpha})}\partial_{\nu}\Phi_{\alpha} - \eta_{\mu \nu}{\cal L}. 
\end{equation}
First, the energy density $\epsilon$ and the pressure $P$ are given by:
\begin{eqnarray}
\label{eq:energy-pressure}
\epsilon &=& \langle T_{00}\rangle  \nonumber \\
&&=\frac{1}{2\pi^2}\sum_{b} \gamma_b \int_0^{\infty} dk k^2\sqrt{k^2 + m_b^{*2}}\,(f_{b}(k,T) + f_{\bar{b}}(k,T)) \nonumber \\
&& +  \frac{1}{2\pi^2} \sum_{l} \gamma_l \int_0^{\infty} dk k^2\sqrt{k^2 + m_l^{2}}\, (f_{l}(k,T) + f_{\bar{l}}(k,T)) \nonumber \\
&& + \frac{1}{2}(m_{\omega}^2 \bar \omega^2+m_{\rho}^2\bar \rho^2+m_{\phi}^2\bar \phi^2+ m_{\sigma}^2\bar\sigma^2+m_{\sigma^{*}}^2\bar \sigma^{{*}^2}) \nonumber \\
&& +\frac{\kappa}{3!}(g_{\sigma N}\bar \sigma)^3 + 
\frac{\lambda}{4!}(g_{\sigma N}\bar \sigma)^4  \nonumber \\
&& + \frac{\zeta}{8}(g_{\omega N}\bar \omega)^4 + 3\Lambda_{\omega}(g_{\rho N}g_{\omega N}\bar \rho \bar \omega)^2 ,\nonumber \\
P &=& \frac{1}{3}\langle T_{jj} \rangle \nonumber \\
&&= \frac{1}{6\pi^2}\sum_{b} \gamma_b \int_0^{\infty} dk \frac{k^4}{\sqrt{k^2 + m_b^{*2}}}{(f_{b}(k,T) + f_{\bar{b}}(k,T))}  \nonumber \\
&&+ \frac{1}{6\pi^2}\sum_{l} \gamma_l \int_0^{\infty} dk \frac{k^4}{\sqrt{k^2 + m_l^{2}}}(f_{b}(k,T) + f_{\bar{b}}(k,T))  \nonumber \\
&&+ \frac{1}{2}(m_{\omega}^2\bar \omega^2+m_{\rho}^2\bar \rho^2+m_{\phi}^2\bar \phi^2-m_{\sigma}^2\bar \sigma^2- m_{\sigma^{*}}^2\bar \sigma^{{*}^2}) \nonumber \\
&&- \frac{\kappa}{3!}(g_{\sigma N}\bar \sigma)^3 -
\frac{\lambda}{4!}(g_{\sigma N}\bar \sigma)^4 \nonumber \\
&& + \frac{1}{24}\zeta(g_{\omega N}\bar \omega)^4 \nonumber \\
&& + \Lambda_{\omega}(g_{\rho N}g_{\omega N}\bar \rho \bar \omega)^2,
\end{eqnarray}
Then, the entropy density $s$ and the free energy density of the system $f$ can be obtained using the following thermodynamic relations
\begin{eqnarray}
\label{eq:entropy-freeenergy}
 s &=& \frac{1}{T}\left(\epsilon + P - \sum_i \mu_i \rho_i \right) , \nonumber \\
 f &=&  \sum_i \mu_i \rho_i - P.
\end{eqnarray}

In the present work, we consider the FSU2H parameterization with nucleons and hyperons of Refs.~\cite{Tolos:2016hhl,Tolos:2017lgv} and extend it to finite temperature. For completeness, we show the mesons masses and nucleon couplings of the FSU2H model in Table~\ref{table:nuclearparam}.
We have, however, improved some of the features of the FSU2H parameterization by, first, modifying the coupling of the $\sigma$ field to the $\Xi$ baryon in order to reproduce a deeper nuclear potential, $U_{\Xi}=-24$ MeV, as recently determined in Ref.~\cite{Friedman:2021rhu} and, secondly, incorporating the $\sigma^*$ field in the Lagrangian density, so as to have a good description of the $\Lambda \Lambda$ bond energy in $\Lambda$ matter \cite{Takahashi:2001nm},\cite{Ahn:2013poa}, while keeping the SU(6) value of the $\phi \Lambda$ coupling. As a result of these changes, also new couplings of $\sigma^* \Sigma$ and $\sigma^* \Xi$ have been introduced, following a strange-quark counting rule. The resulting model is denoted as FSU2H$^*$. In Table~\ref{table:hypparam} we summarize the hyperonic couplings to the meson fields in terms of their ratios to the corresponding couplings of nucleons: $R_{i Y}=g_{iY}/g_{iN}$ for $i={(\sigma,\omega,\rho)}$, and $R_{\sigma^* Y}=g_{\sigma^* Y}/g_{\sigma N}$ and $R_{\phi Y}=g_{\phi Y}/g_{\omega N}$, since $g_{\sigma^* N}=0$ and $g_{\phi N}=0$ due to the OZI rule. Note that the isospin operator $I_{3i}$ appearing in
Eq.~(\ref{eq:mueff}) converts the 1:1 relation between the $g_{\rho \Sigma}$ and $g_{\rho N}$ couplings seen in Table~\ref{table:hypparam} into the proper 2:1 ratio expected by the symmetries assumed.

\begin{table}
\centering
\caption{The ratios of the couplings of hyperons to mesons with respect to the nucleonic ones.}
\label{tab_2_2}       
\begin{tabular}{cccccc}
\hline\noalign{\smallskip}
$Y$ & $R_{\sigma Y}$ &$R_{\omega Y} $ & $R_{\rho Y}$& $R_{\sigma^{*}Y}$ & $R_{\phi Y}$ \\
\noalign{\smallskip}\hline\noalign{\smallskip}
$\Lambda$ & $0.6113$ & $2/3$ & $0$ & $0.2812$ & $-\sqrt{2}/3$  \\
$\Sigma$ & $0.4673$ & $2/3$ & $1$ & $0.2812$ & $-\sqrt{2}/3$  \\
$\Xi$ & $0.3305$ & $1/3$ & $1$ & $0.5624$ & $-2\sqrt{2}/3$  \\
\hline\noalign{\smallskip}
\end{tabular}
\label{table:hypparam}
\end{table}

\section{Composition and Equation of state}
 \label{sec:2}

Due to the evolving temperature, the composition and the thermodynamical properties of neutron star matter may change significantly during the neutron star formation or in other processes it might be involved with, such as neutron star mergers. In order to explore every situation of interest, we will consider the core of the star at two different and extreme temperatures, $T = 5$ MeV and $ T = 50$ MeV,  and different lepton fractions $Y_l = 0.2, 0.4$, as well as the case where neutrinos have already diffused out from the star, so the lepton number is no longer conserved. We note that the muon lepton number in all calculations of neutrino trapped matter is fixed to $Y_{\mu}=0$. The hot and lepton rich matter of the star corresponds to the earlier evolution stages, while neutrino-free case corresponds to the later times \cite{Pons1999EvolutionStars,OConnor2018ExploringSupernovae,2017RSPTA.37560271C,Sumiyoshi2007DynamicsDependence,Fischer:2008rh}.

\subsection{Composition}
\begin{figure}
    \centering
    \includegraphics[width=0.48\textwidth]{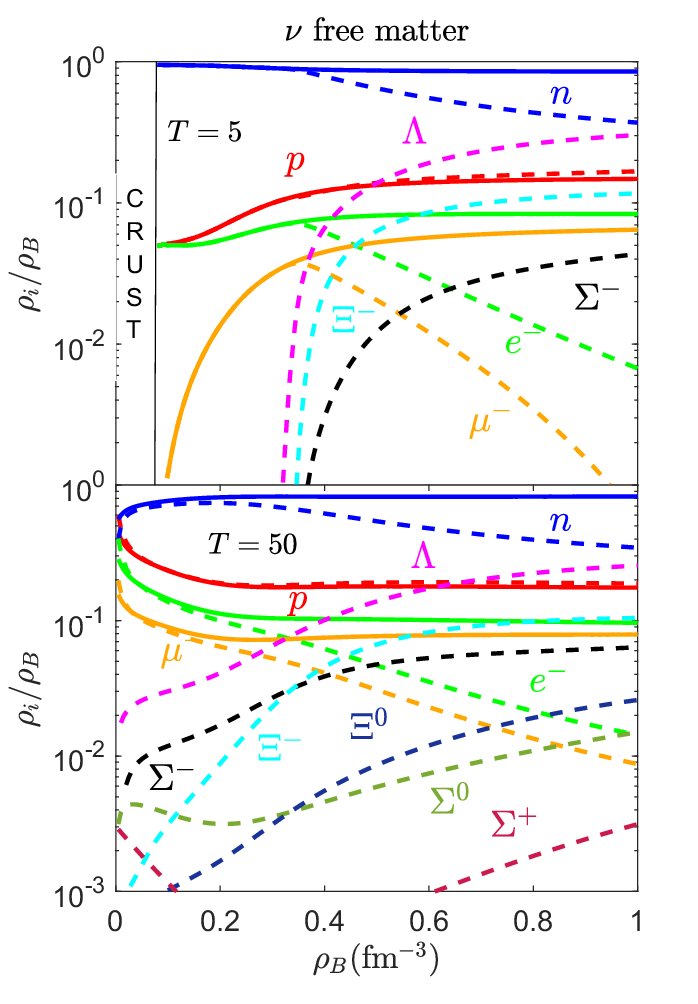}
    \caption{The composition of the neutron star core for neutrino free matter, without (solid lines) and with (dashed lines) hyperons, for $T=5$ MeV (upper panel) and $T=50$ MeV (lower panel).}
    \label{fig:1,2}
\end{figure}
In this subsection we show the composition of the star for the above-mentioned different cases. Note that, due to the inclusion of the antiparticle contribution, the density of each species corresponds to the difference between the particle abundance and the corresponding antiparticle one. In 
Fig.~\ref{fig:1,2} the composition of a neutrino free neutron star core is shown for temperatures $T=5$ MeV (upper panel) and $T=50$ MeV (lower panel). The different colors in the figures represent the relative abundance of the particles considered as function of the baryonic density $\rho_B$. Solid lines correspond to the case where hyperons are not considered, while dashed lines take the appearance of hyperons into account in the core of the star. The composition pattern at temperatures as low as $T=5$ MeV is essentially the same as that for cold neutron star matter \cite{Tolos:2016hhl,Tolos:2017lgv}. We can see that when hyperons are not included in the core, the particle abundances slowly vary with respect to the density. At higher temperature, matter tends to be slightly more isospin symmetric. The reason is that temperature lowers the chemical potentials, specially for the lighter particles, such as the leptons, thereby making the conversion $n \to p + e^-$ more efficient. This is also the reason for the presence of muons already from the beginning of the core. 
The composition pattern changes significantly with temperature when hyperons are included. At any temperature we can notice that, as soon as hyperons appear, a deleptonization process takes place. Negative hyperons ($\Sigma^{-}, \Xi^{-}$) now enter the conserving charge neutrality condition and they can replace the ``expensive" leptons (in terms of energy density and pressure), which are slowly vanishing from the core as density increases. It is also noticeable that the inclusion of hyperons lowers the neutron abundance. At high enough density neutrons are highly energetic, hence they can be replaced by $\Lambda$ baryons, a species which at $3\rho_0$  is already more abundant than the proton. We observe that the onset density for the appearance of hyperons, at low temperature, corresponds essentially to that at $T=0$, namely around $2\rho_0$ \cite{Baldo:1998hd,Vidana2000Hyperon-hyperonMatter}. However, at high temperatures, the hyperons are present even at the beginning of the core at low densities, and all hyperons of the baryon octet can be found in the core of the star. Still, due to their charge and mass, the most important ones are $\Lambda, \Sigma^{-}$ and $\Xi^{-}$. We note that, although being lighter than the $\Xi^{-}$, the $\Sigma^{-}$ is assumed to feel a nuclear repulsive potential \cite{Noumi:2001tx,Saha:2004ha,Kohno:2006iq,HARADA2006206}, which makes this species to be less abundant than the $\Xi^{-}$ at densities larger than $2\rho_0$, in qualitative agreement with with other models that make the same assumption \cite{Sedrakian:2021qjw,Raduta:2020fdn}.

\begin{figure}
    \centering
    \includegraphics[width=0.48\textwidth]{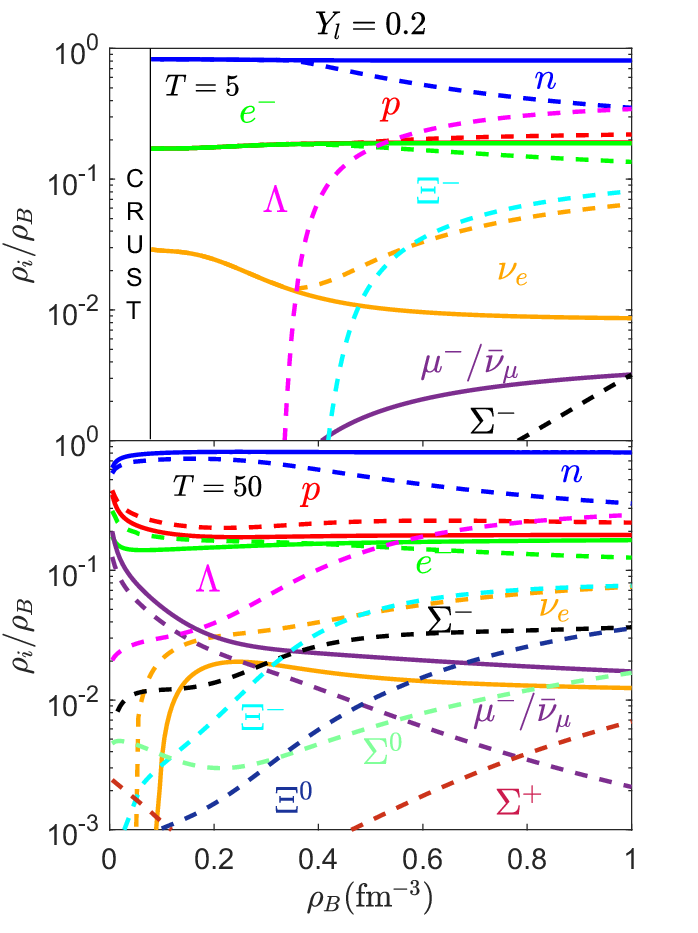}
    \caption{The composition of the neutron star core for neutrino trapped matter with lepton number $Y_l = 0.2$, without (solid lines) and with (dashed lines) hyperons, for $T=5$ MeV (upper panel) and $T=50$ MeV (lower panel).}
    \label{fig:3,4}
\end{figure}

We now discuss the cases that are important at the early evolution times of the star, when neutrinos are trapped in its core. Again, we consider two different temperatures, $T=5$ MeV and $T=50$ MeV, but now at two different lepton fractions, $Y_l = 0.2$ (Fig.~\ref{fig:3,4}) and $Y_l = 0.4$ (Fig.~\ref{fig:5,6}). As for the neutrino-free case, we observe that when hyperons are not included the abundance of each species does not change significantly with increasing density. Comparing the composition patterns, one can observe that a higher lepton fraction implies an increased electron fraction, which brings the fraction of the protons closer to that of neutrons, making the matter more isospin symmetric. This conclusion does not depend on the temperature of the core. The inclusion of hyperons gives rise to significantly different particle abundances at both temperatures considered. Similarly as for the neutrino-free case, at sufficiently high temperature, hyperons are present all over the core, increasing their abundances with the lepton fraction and all of them reaching significant values (above 0.01$\rho_B$) at high densities. Note that higher lepton fractions imply a higher chemical potential of the protons, which enables for the positively charged baryon $\Sigma^{+}$ to have a significant abundance too. A closer inspection of the bottom panel of Fig~\ref{fig:5,6} reveals that, at the highest density considered $\rho_B = 1.0$ fm$^{-3}$, the partial densities of the baryons from the same isospin family tend to converge to a special isospin degeneracy point, as was pointed out in \cite{Sedrakian:2021qjw}. This implies that, at that particular density, the $\rho$ meson field vanishes. However, the exact density at which this degeneracy point exists is model dependent. 
\begin{figure}
    \centering
    \includegraphics[width=0.48\textwidth]{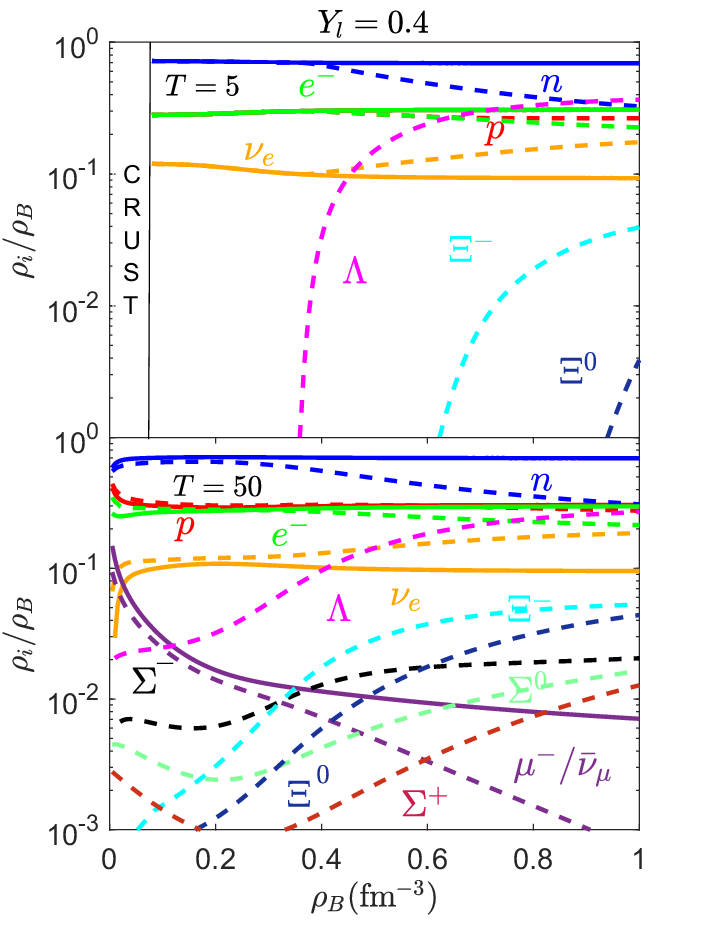}
    \caption{Same as Fig. \ref{fig:3,4} but for $Y_l = 0.4$.}
    \label{fig:5,6}
\end{figure}

It is also interesting to pay attention to the relative abundance of neutrinos in Figs.~\ref{fig:3,4} and \ref{fig:5,6}. As we have already commented, the appearance of negative hyperons lowers the abundance of the electrons. However, as the electron lepton number is conserved, the abundance of neutrinos increases significantly, specially at high densities. This observation is important because the transport of energy driven by neutrinos plays a crucial role in fueling supernovae explosions, as well as in the dynamics of neutron stars mergers, hence having a strong influence in the hydrodynamic simulations of such phenomena \cite{2017hsn..book.1575J,2022EPJA...58...99C}.
We also see that the presence of electron neutrinos hinders the appearance of muons. Note that, due to the conservation of the lepton numbers, the muons can be created by the reaction
$e^{-} = \mu^{-} + \nu_{e}+\bar{\nu}_{\mu}$,
which gives the relation between chemical potentials of the species given with Equation \ref{eq:beta_chemical_potentials}.
Hence, an increase of the electron neutrino abundance implies a decrease of the two equally abundant muonic species ($\rho_{\mu} = \rho_{{\bar \nu}_{\mu}}$), which explains the rare abundance of the muons, specially in the case in which hyperons are present in the neutron star core.



\subsection{Equation of state}
\begin{figure*}
    \centering
        \includegraphics[width=1 \textwidth]{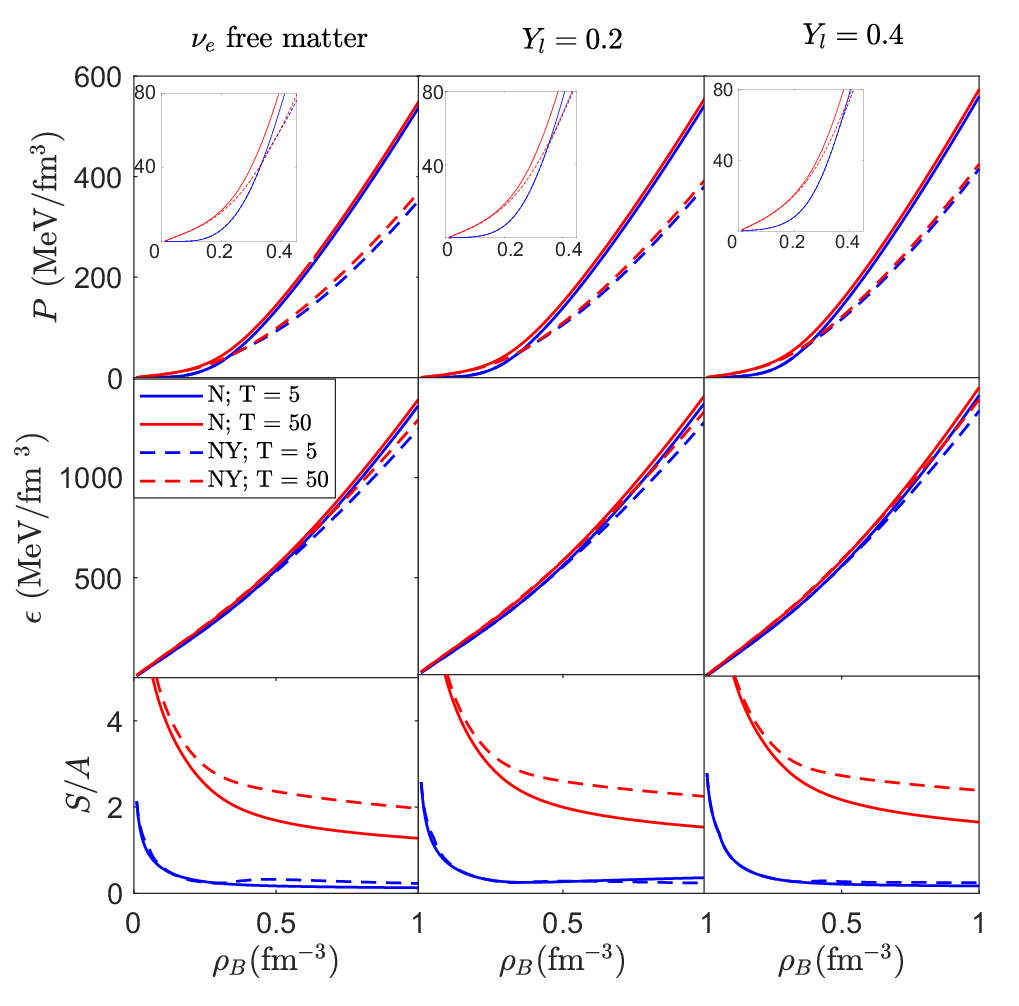}
         \setlength{\unitlength}{0.01\linewidth}
    \caption{The pressure $P$ , energy density $\epsilon$ and entropy per baryon $S/A$ as a function of the baryonic density $\rho_B$ for temperatures $T=5$ MeV and $T=50$ MeV, without (solid lines) and with (dashed lines) hyperons. The different columns correspond to the neutrino free case (left column), lepton fraction $Y_l=0.2$ (middle column) and $Y_l=0.4$ (right column). }

\label{fig:EoS}
\end{figure*}

As it has been pointed out in the previous subsection, the composition of the star is sensitive to the temperature of the core and to the  species taken into account. As a result, the EoS for different scenarios varies depending on the composition. In Fig.~\ref{fig:EoS} we present the pressure (upper panels), energy density (midle panels) and entropy per baryon (bottom panels) for both neutrino free and neutrino trapped matter ($Y_l = 0.2$ and $Y_l = 0.4$) at $T = 5$ MeV (blue curves) and $T = 50$ MeV (red curves). As before, the solid lines (N curves) correspond to calculations where hyperons are not taken into account, while the dashed lines (NY curves) are obtained considering hyperons too.


In all different cases we can see that hyperons produce a significant softening of the EoS, a very well known fact since the first calculations including hyperons \cite{Pandharipande:1971up,Balberg1997AnStrangeness,Vidana2000Hyperon-hyperonMatter,Vidana:2002rg}. At low temperatures, the NY $P(\rho_B)$ curve drastically changes its slope when hyperons start to appear in the core, as can be seen more clearly in the figure insets. On the contrary, at higher temperatures, the slope changes more smoothly, since hyperons are already present in the core of the star at low densities. It is also clear that  temperature effects are more important when hyperons are present in the core of the star. An increment of the number of species in the core produces more states available for the particles to occupy, making the matter less degenerate, and hence, more affected by temperature. This is specially important at the low densities of the core, when the degeneracy is the lowest.

Although not very clear from the plots in Fig.~\ref{fig:EoS}, similar conclusions can be drawn about the effects of temperature and hyperons on the $\epsilon (\rho_B)$ dependence.
It is also worth pointing out that the EoS becomes stiffer as one increases the lepton fraction in the neutrino trapped case. In the previous section we saw that, in the absence of neutrinos, the appearance of hyperons implied a deleptonization of matter, which in turns makes the EoS softer. In contrast, when the lepton fraction is fixed, this mechanism is not possible since the total density of the leptons remains constant. As a consequence, the EoS in the neutrino trapped case is stiffer than that in which neutrinos have diffused out of the core of the star. 

\begin{figure}
    \centering
    \includegraphics[width=0.48 \textwidth]{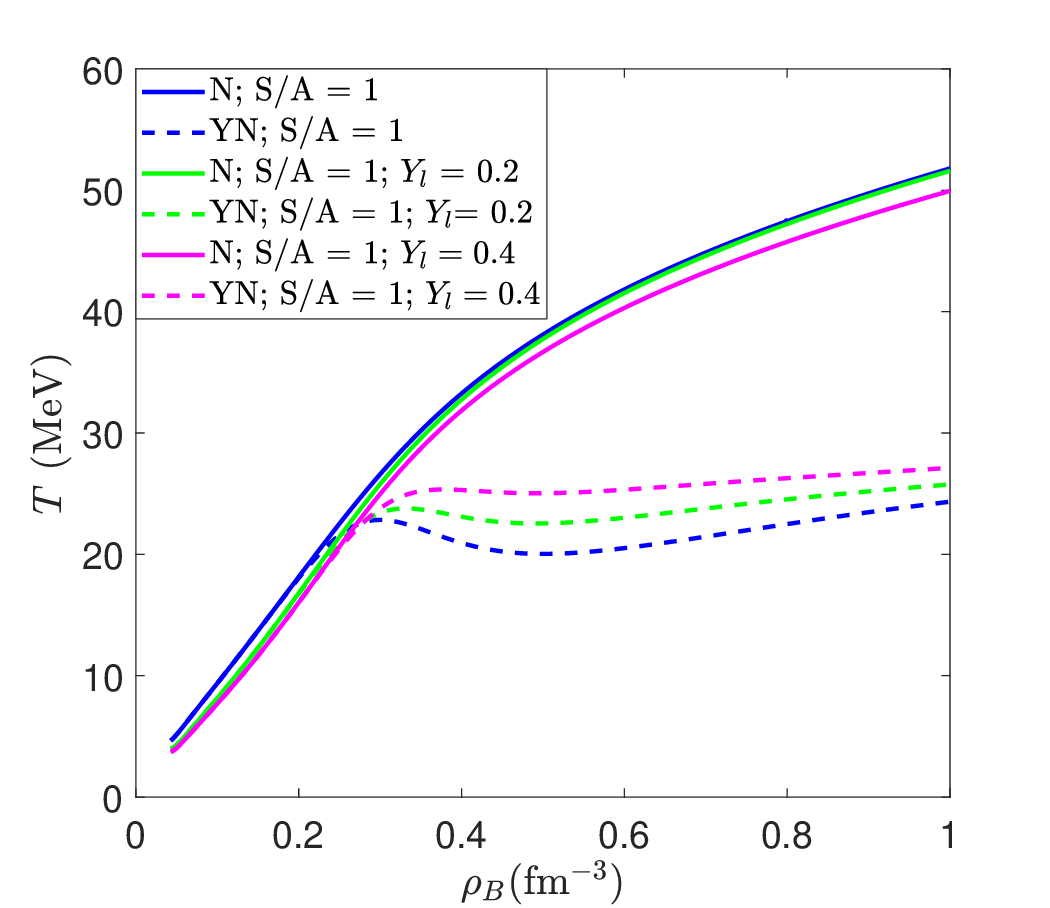}
    \caption{Temperature profile of $\beta-$ stable matter at fixed entropy per baryon $S/A = 1$. Solid lines correspond to the cases where hyperons are not considered in the core of the star, while dashed lines represent cases where they are taken into account.}
    \label{fig:t_profile}
\end{figure}

The behavior of the entropy per particle $S/A$ as a function of the baryon density confirms the aforementioned conclusions. As expected, the entropy per particle increases with temperature but the effect is larger when hyperons are included.  As the density increases, the entropy per particle in the nucleonic case decreases. Hyperons slow down this decrease and even change the monotonous behaviour of the curve in a certain density range in some cases (see the neutrino-free case at low temperatures). Many of the properties of the stars, such as the mass-radius relation, are usually computed assuming that an isentropic profile and, therefore, a change in the trend of the entropy can have an impact on these properties.
We also observe that the presence of hyperons can serve as temperature control mechanism, making the temperature almost constant throughout the most dense parts of the core.
This is better seen in Fig.~\ref{fig:t_profile}, where we show the temperature profile of $\beta{-}$ stable matter at constant entropy per baryon ($S/A = 1$). It is clear that when hyperons are included, the temperature tends to a constant value as we increase the density, thus corroborating the aforementioned statement that hyperons serve as a temperature control mechanism. The effect is stronger in the neutrino free case when the abundance of hyperons is not hindered by the conservation of the lepton number. This conclusion is also confirmed with other models too, in particular, for the calculation performed at constant entropy per baryon in Ref.~\cite{Raduta:2020fdn} employing the DDME2Y model. 


\subsection{Thermal index in $\beta$-stable matter}
We will close the finite-temperature treatment of the EoS with a calculation of the thermal index, which is defined as:
\begin{equation}
\centering
    \Gamma(\rho_B,T) \equiv 1 + \frac{P_{\rm th}}{\epsilon_{\rm th}} \ ,
    \label{eq:gamma}
\end{equation}
where the thermal pressure $P_{\rm th}$ and energy density $\epsilon_{\rm th}$ are given by:
\begin{eqnarray}
P_{\rm th}&=& P(\rho_B,T)-P(\rho_B,T=0)  \nonumber \\
\epsilon_{\rm th}&=& \epsilon(\rho_B,T)-\epsilon(\rho_B,T=0)  \ ,
\label{eq:thermal}
\end{eqnarray}
with $P(\rho_B,T=0)$ and $\epsilon(\rho_B,T=0)$ being the pressure and the energy density in cold matter, respectively.
The above definition of the thermal index is done in analogy with the $\Gamma$ law relating the pressure and energy density of a diluted ideal gas, where $\Gamma$ is the constant adiabatic index with a value of $5/3 (4/3)$ in the case of non-relativistic (ultra-relativistic) fermionic systems. The pressure, $P(\rho_B,T)$, and energy density, $\epsilon(\rho_B,T$), of neutron-star matter obviously deviate from the ideal gas behavior due to the strong correlations among the nucleons, and eventually hyperons, present in the core. For this reason, it is expected that the thermal pressure and energy density given by Eq.~(\ref{eq:thermal}) produce a temperature and density dependent thermal index.

Assuming a mild dependence, the use of Eq.~(\ref{eq:gamma}) with a constant value of the thermal index $\Gamma$ has been proved to be very practical in complicated and numerically costly merger simulations. Usually, the conditions in these simulations require the finite temperature EoS within a range of $10-100$ MeV. In order to describe a large variety of physical phenomena while reducing significantly the computational costs, many groups \cite{Hotokezaka:2013iia,Endrizzi:2018uwl,Camelio:2020mdi} rely on the pressure and energy density of cold matter, for which there are many models in the literature, to which they add the thermal correction, $P_{\rm th}$, which is obtained from the relation established by Eq.~(\ref{eq:gamma}), treating $\epsilon_{\rm th}$ as an independent variable and taking a constant value for $\Gamma$ of usually between 1.5 and 2.0. With this procedure, an equation of state $P = P (\epsilon)$ at finite temperature is obtained. While this approach is fast, and reliable if the matter can be considered as an ideal fluid, it can be inaccurate at the conditions of neutron stars mergers \cite{Bauswein:2010dn,Raithel2021RealisticSimulations}.

\begin{figure}
    \centering
    \includegraphics[width=0.48 \textwidth]{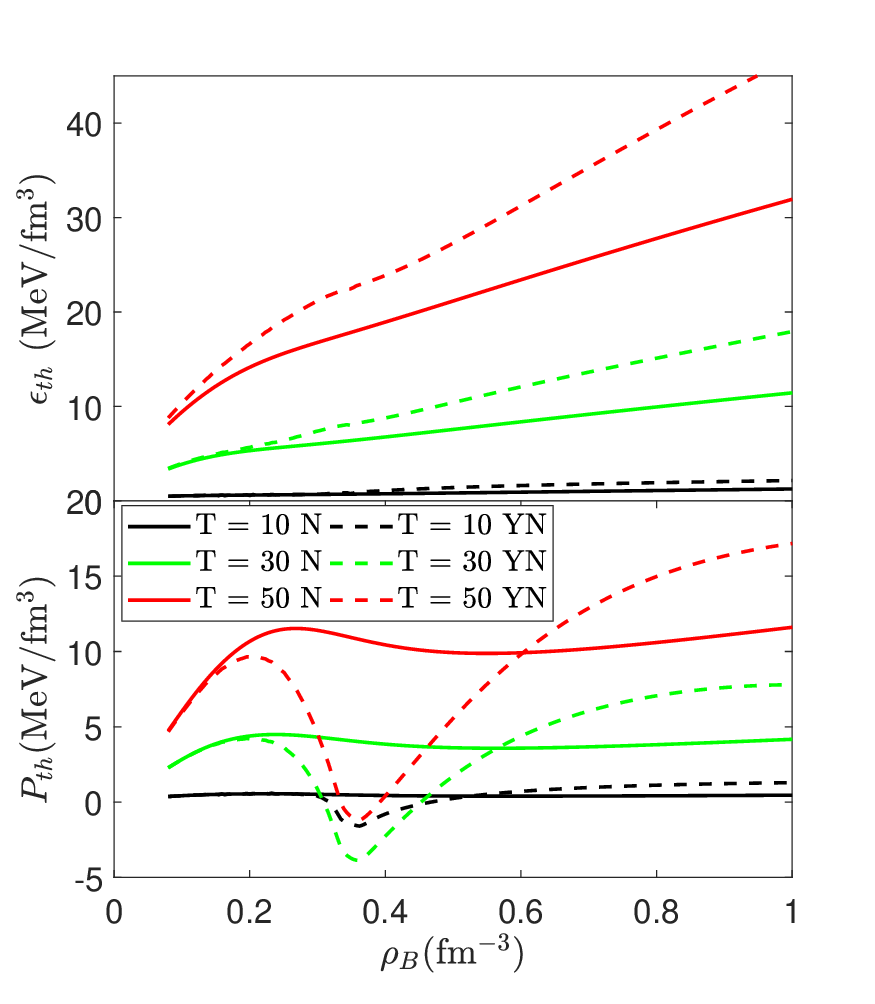}
    \caption{The thermal energy $\epsilon_{th}$ and the thermal pressure $P_{th}$ as a function of the baryonic density $\rho_B$ for various temperatures (in MeV units) and for the neutrino free case. The solid lines correspond to matter without hyperons, whereas the dashed ones are for matter with hyperons.}
    \label{fig:thermal_energy_pressure}
\end{figure}
 
\begin{figure}
    \centering
    \includegraphics[width=0.48 \textwidth]{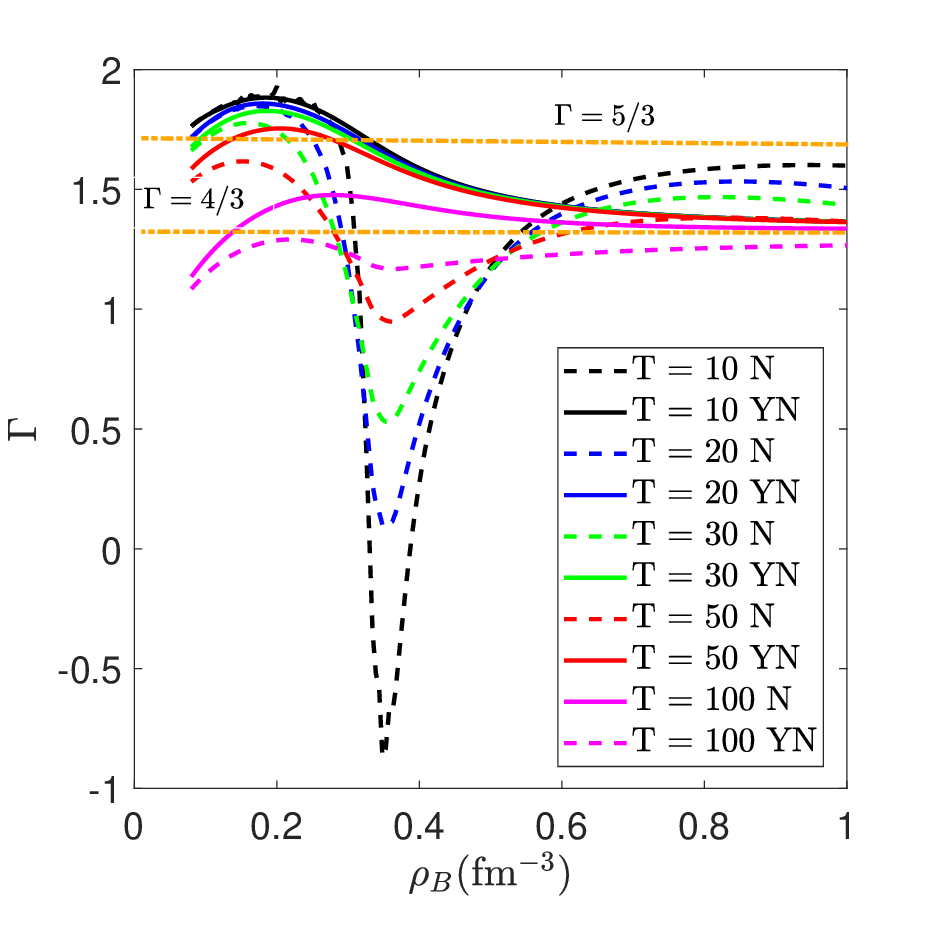}
    \caption{The thermal index $\Gamma$ as a function of the baryonic density $\rho_B$ for various temperatures (in MeV units) and for the neutrino free case,  without (solid lines) and with (dashed lines) hyperons. The dashed doted orange lines $\Gamma = 5/3$ and $\Gamma = 4/3$ correspond to the thermal index of a non-relativistic and ultra-relativistic diluted ideal gas, respectively.}
    \label{fig:thermal_index}
\end{figure}

In this section, we obtain the thermal pressure and energy density from Eq.~(\ref{eq:thermal}), using our equations of state for zero and finite temperature, and investigate how the thermal index, defined in Eq.~(\ref{eq:gamma}), evolves with the temperature and density, paying a special attention to the role played by hyperons.
We will focus on the neutrino free case. In Fig.~\ref{fig:thermal_energy_pressure} we show $P_{\rm th}$ and $\epsilon_{\rm th}$ for three different temperatures as functions of baryon density for nucleonic only matter (solid lines) and hyperonic matter (dashed lines). We see that the inclusion of hyperons does not change the monotonously increasing behaviour of $\epsilon_{\rm th}$ with density. However, their appearance has a much more evident effect on $P_{\rm th}$. At densities in which hyperons achieve a significant abundance (which is correlated to the threshold density for the appearance of hyperons at $T=0$, i.e., $\rho_B = 0,35\ \ \text{fm}^{-3}$), the thermal pressure experiences a sizable drop, which is deeper at higher temperatures. The value of $P_{\rm th}$ quickly recovers at larger densities and ends up overshooting the value obtained in the absence of hyperons. This complex behaviour of $P_{\rm th}$ heavily influences the thermal index, which is displayed in Fig.~\ref{fig:thermal_index} for several temperatures. The hyperonic results (dashed lines) show the sudden drop around $\rho_B=0.35 $ fm$^{-3}$, which is the density where the hyperon abundance starts to be significant, even at low temperatures. One can also point out that, with increasing temperature, the drop becomes shallower, due to the larger increase of $\epsilon_{\rm th}$, and wider, because hyperons are already present at the lower densities of core. 
We observe that, at higher densities, the thermal indices for the hyperonic equations of state are higher than the corresponding nucleonic ones. 
Note that, when nucleons are the only hadrons present in the core of the star, they are more abundant and can achieve the degenerate ultra-relativistic regime, for which the thermal index has the value of $\Gamma = \frac{4}{3}$, already at densities of around $4-5 \rho_0$. We note that the effective mass of nucleons at these high densities is significantly smaller than their bare mass. This behavior is rather independent of the temperature, as expected for a degenerate case. When hyperons are present, they fill lower momentum states and, at the same time, the higher momentum nucleonic states are emptied, both effects contributing to move the thermal index away from the ultra-relativistic limit, as can be seen from the dashed lines in Fig.~\ref{fig:thermal_index}. Note, however, that there is some temperature dependence in this case. This is tied to the fact that the neutrons are less abundant at lower temperatures and densities, reducing the amount of high momentum components. 

\begin{figure}
    \centering
    \includegraphics[width=0.46 \textwidth]{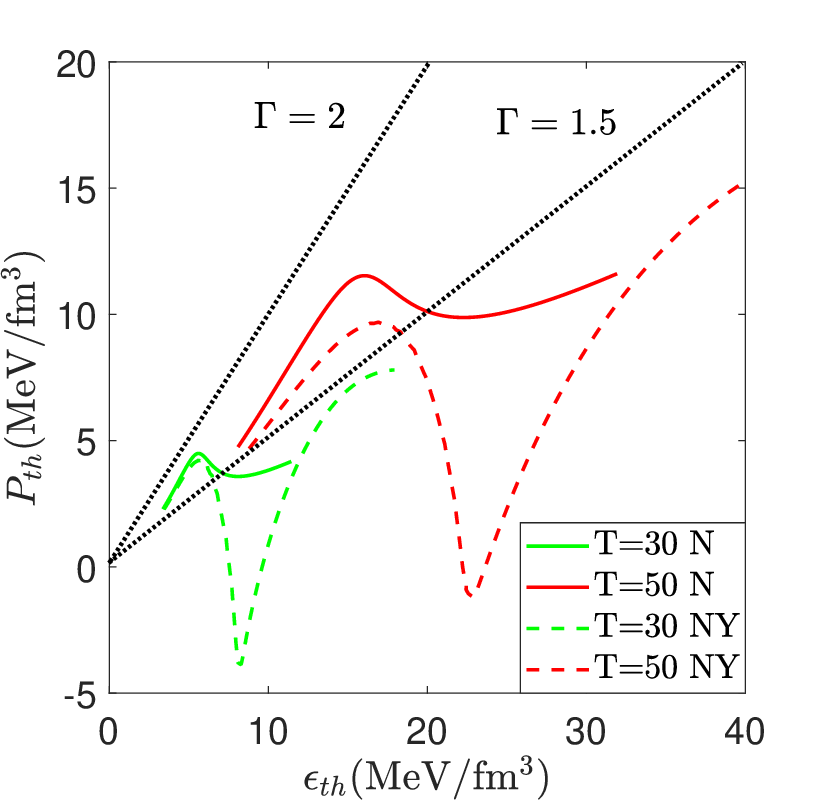}
    \caption{The thermal pressure $P_{th}$ as a function of the thermal energy density $\epsilon_{th}$ for different temperatures (in MeV units), without (solid lines) and with hyperons (dashed lines). The dotted lines represent $P_{th}(\epsilon_{th})$ for a constant thermal index of $\Gamma=2$ (upper line) and $\Gamma=1.5$ (lower line). }
    \label{fig:p_thermal_e_thermal}
\end{figure}

A better representation of the inaccuracy that may be inflicted when thermal effects are computed with a constant thermal index is given in Fig.~\ref{fig:p_thermal_e_thermal}, where we represent $P_{\rm th}$ versus $\epsilon_{\rm th}$, employing our nucleonic (solid lines) and hyperonic (dashed lines) equations of state at different temperatures and densities. Each line covers our results for the whole range of density values considered in the present work ($0.08-1.0$ fm$^{-3}$) at a constant temperature. Note that, although only two temperatures have been included in the figure, the consideration of a whole set of different temperatures would generate a family of nucleonic curves and a family of hyperonic curves covering certain regions of the $P_{\rm th}-\epsilon_{\rm th}$ plane. The dotted black lines represent two typical extreme linear behaviors assumed merger simulations.
The results of Fig.~\ref{fig:p_thermal_e_thermal} clearly show that the $P_{\rm th}(\epsilon_{\rm th})$ behavior is much more complex than the linear assumption. For a given value of $\epsilon_{\rm th}$, the corresponding value of $P_{\rm th}$ at different temperatures is noticeably distinct, contrary to the fixed thermal correction at all temperatures that a constant thermal index would imply. For instance, at $\epsilon_{\rm th}=10$ ${\rm MeV/fm^3}$, the value of $P_{\rm th}$ at $T=30$ MeV in the nucleonic case is 3 ${\rm MeV/fm^3}$ lower than that at $T=50$ MeV. In the hyperonic case, this difference is twice as large. One could still argue that a constant index of $\Gamma=1.5$ would reproduce, on average, the behavior of the solid lines, obtained with the nucleonic equation of state. However, the presence of hyperons would invalidate the use of such a constant value of $\Gamma$ and even of any other constant value, given the larger region in the $P_{\rm th}-\epsilon_{\rm th}$ plane covered by the hyperonic calculations.

\begin{figure}
    \centering
    \includegraphics[width=0.48 \textwidth]{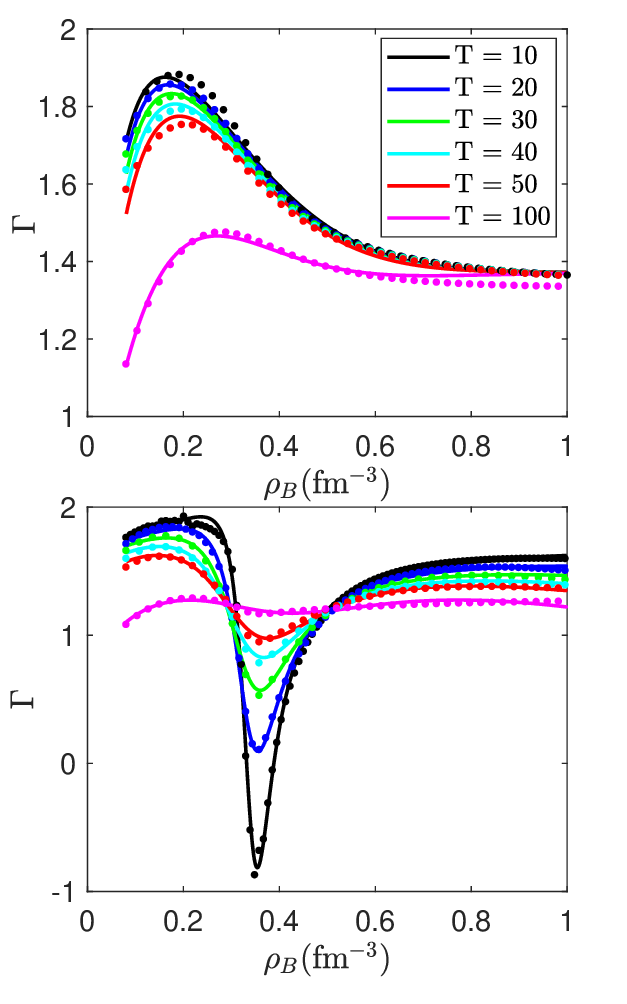}
    \caption{$\Gamma = \Gamma(\rho_B,T) $ fit as a function of the baryonic density $\rho_B$ for the neutrino free case, without (upper panel) and with hyperons (lower panel). Dots are representing the exact calculations, while the solid lines are the results from the fits. The temperatures shown are in MeV units.}
     \label{fig:thermal_index_fit}
\end{figure}

In order to use the concept of thermal index for $\beta$ stable matter in neutrino free calculations, while, at the same time, reproducing accurately our finite temperature results, we provide a parameterization of $\Gamma$ for the nucleonic and hyperonic cases, covering a range of temperatures from $T=10$ MeV to $T=100$ MeV and a range of densities of from $\rho_B=0.08$ to $\rho_B=1.0$ fm$^{-3}$. The behaviour of the thermal index when hyperons are included in the core of the star is reproduced by the following functional form:
\begin{equation}
\begin{aligned}
\label{Eq_3_1_1}
\Gamma (\rho_B,T) &=&\frac{a(T)\rho_B+ b(T)}{(\rho_B-{\rho_B}_c)^2+c(T)\rho_B +d(T)}  \\
& & + ~e(T)\rho_B +   f(T)\sqrt{\rho_B}+g(T) \ ,
\end{aligned}
\end{equation}
with ${\rho_B}_c = 0.3715$ fm$^{-3}$, and where the temperature dependent functions are polynomials:
\begin{equation}
\begin{aligned}
\label{Eq_3_1_2}
&a(T) = a_1T+a_2 && \\
&b(T) = b_1T+b_2 && \\
&c(T) = c_1T^2+c_2T+c_3 && \\
&d(T) = d_1T^2+d_2T+d_3 && \\
&e(T) = e_1T^2+e_2T+e_3 && \\
&f(T) = f_1T+f_2 && \\
&g(T) = g_1T^2+g_2T+g_3 \ .
\end{aligned}
\end{equation}

A simpler form has been found to reproduce our results when a purely nucleonic core is considered:
\begin{equation}
\Gamma (\rho_B,T) = a_1 + b_1\, \text{e}^{-c(T)\rho_B^2} + d_1\, \text{e}^{-e(T)\rho_B},
\end{equation}
where the functions $c(T)$ and $e(T)$ have a linear dependence with $T$:
\begin{equation}
\begin{aligned}
\label{Eq_3_1_4}
&c(T) = c_1T+c_2 &&\\
&e(T) = e_1T+e_2 \ .
\end{aligned}
\end{equation}

The
comparison of the fit and the exact calculations is given in Fig. \ref{fig:thermal_index_fit}. The values of the parameters for the nucleonic thermal index are listed in Table \ref{tab_3_2} and those for the hyperonic one in Table \ref{tab_3_1}. We note that the units of every coefficient are such that the thermal index is a dimensionless quantity. The temperature has units of MeV, while the baryon density is given in $\text{fm}^{-3}$.

\begin{table*}
\centering
\begin{center}
\begin{tabular}{cccccccc}
\hline\noalign{\smallskip}
$i$  & $a_i$ & $b_i$ & $c_i$ & $d_i$ & $e_i$   \\
\noalign{\smallskip}\hline\noalign{\smallskip}
1 &$1.37 $& $6.89\cdot10^{-1}$ & $1.69\cdot 10^{-2}$ & $-1.47$ & $-1.41 \cdot 10^{-1}$   \\
\noalign{\smallskip}\hline
2 &$/$ & $/$& $6.90$ & $/$ & $20.29$   \\
\hline\noalign{\smallskip}
\end{tabular}
\end{center}
\caption{The thermal index coefficients when a nucleonic core is considered.}
\label{tab_3_2}  
\end{table*}

\begin{table*}
\centering
\begin{tabular}{cccccccccc}
\hline\noalign{\smallskip}
$i$  & $a_i$ & $b_i$ & $c_i$ & $d_i$ & $e_i$ & $f_i$ &$g_i$ \\
 
\noalign{\smallskip}\hline\noalign{\smallskip}
1 &$-2.02\cdot10^{-4}$ & $-1.26\cdot10^{-4}$  &$ -1.71\cdot10^{-5}$ & $1.24\cdot10^{-5}$ & $1.11\cdot10^{-4}$ & $3.21\cdot10^{-2}$ & $-6.64\cdot10^{-5}$  \\
\noalign{\smallskip}\hline
2 &$-6.14\cdot10^{-2}$ & $1.98\cdot10^{-2}$ & $1.83\cdot10^{-3}$ & $-6.22\cdot10^{-4}$ & $-3.63\cdot10^{-2}$ & $2.60\cdot10^{-1}$ & $-5.13\cdot10^{-3}$  \\
\noalign{\smallskip}\hline
3 & / & / & $1.56\cdot10^{-2}$ & $-4.83\cdot10^{-3}$ & $7.15\cdot10^{-2}$ & / & $1.48$\\
\hline\noalign{\smallskip}
\end{tabular}
\caption{The thermal index coefficients when a hyperonic core is considered.}
\label{tab_3_1}       
\end{table*}

\begin{figure*}
    \centering
    \includegraphics[width=1.0 \textwidth]{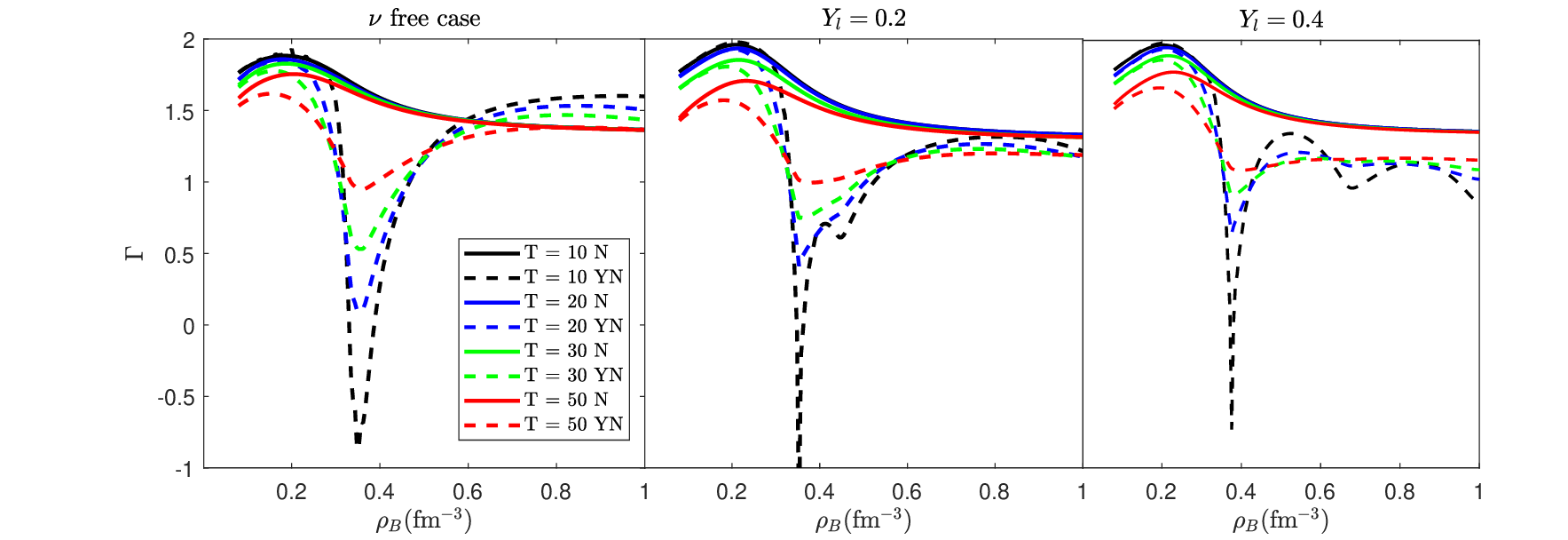}
    \caption{The thermal index $\Gamma$ as a function of the baryonic density $\rho_B$ for various temperatures (in MeV units), and for neutrino free case and for the different lepton fractions, without (solid lines) and with (dashed lines) hyperons.}
    \label{fig:thermal_index_different_cases}
\end{figure*}

For the sake of completeness, we also show the thermal index in the neutrino-trapped case, for two different lepton fractions, $Y_l = 0.2$ and $Y_l = 0.4$, in Fig.~\ref{fig:thermal_index_different_cases}. The behaviour of the curves can be explained in an analogous way as for the neutrino free case. Every drop of the hyperonic curves corresponds to an appearance of a new hyperon species into the star core. This is particularly visible at the lowest shown temperature of $T=10$ MeV, because the onset of hyperons is more abrupt. Note that, in the neutrino-free $T=10$ MeV case, three hyperonic species appear at a very similar density (see Fig.~\ref{fig:1,2}) and this is visualized as a single drop in the corresponding thermal index curve. At higher temperatures the hyperons are already present at the lower densities in the core, so the thermal index has a smoother behaviour. We can also see that at higher lepton fractions ($Y_l = 0.4$), as matter is more isospin symmetric, the chemical potential of the neutron is lower than in the case of neutrino free case, hence limiting the abundance of hyperons. Consequently, their contribution to the total energy density and total pressure is smaller, so the drop of the thermal index is more moderate.

\section{Summary}
\label{sec:3}

We have constructed a model for the core of neutron stars, named FSU2H$^*$, by improving the hyperonic FSU2H scheme of Refs.~\cite{Tolos:2016hhl,Tolos:2017lgv}. On the one hand, we have modified the coupling of the $\sigma$ field to the $\Xi$ baryon so as to reproduce a more attractive value of the $\Xi$ nuclear potential reported recently. On the other hand, we have incorporated the $\sigma^*$ field in order to have a better description of the $\Lambda \Lambda$ bond energy in $\Lambda$ matter.

We have also extended the new FSU2H$^*$ model to include the effects of finite temperature. The correct description of the hot inner core of neutron stars is of fundamental importance for understanding different astrophysical phenomena, such as supernovae explosions and neutron star mergers. We have investigated the composition and EoS of neutron star matter with and without hyperons. We have observed the thermal corrections to have a strong influence on the composition of the inner core, being clearly manifest when hyperons are considered. Whereas only the $\Lambda, \Xi^{-}$ and $\Sigma^{-}$ hyperons are present at low temperatures, all baryons from the baryonic octet can be found all over the core at large temperatures. In particular, when neutrinos are trapped and the lepton fraction is high, a noticeable abundance of $\Sigma^{+}$ can be seen. Furthermore, when hyperons are present, the charged leptonic fractions decrease  and, due to lepton number conservation in the neutrino trapped cases,  the neutrino population increases significantly.

The temperature effects on the thermodynamical properties (pressure, energy density and entropy per baryon) of neutron star matter have also been found to be substantial, specially when hyperons are present, since they lead to a less degenerate core. These effects have been analyzed in terms of the thermal index, which has been observed to depend non-negligibly on temperature and density. For nucleonic matter, the variations of the thermal index with temperature are mild, specially for low leptonic fractions. The nucleonic thermal index shows a stronger but still moderate dependence with density, acquiring values in the range $1.1-2$ over all core densities and lepton fractions explored, and reaching the ultra-relativistic limit of 4/3 already from $\sim 4 \rho_0$ onwards.  
The appearance of hyperons produces a sudden drop in the thermal index, which is quite sizable at low and moderate temperatures ($T<20$ MeV) and can even lead to negative values of $\Gamma$, as the pressure at finite temperature could be lower than that at zero temperature for those densities where hyperons become abundant. After each drop associated to the appearance of a new hyperon species, the thermal index quickly recovers as density increases, and, towards large densities,  it tends to a rather constant value in the range $1.0-1.5$, depending on the lepton fraction, but not converging to the ultra-relativistic value as in the nucleonic case. 

We finally point out that our results for the thermal index can have a strong impact on the simulations of supernovae and neutron stars mergers, that are usually performed taking a constant value throughout the different parts of the core. Therefore, in order to make our calculations easily accessible for such simulations, we have provided analytic parameterizations of the thermal index that reproduce our thermal nucleonic and hyperonic equations of state with high accuracy.


\section*{Acknowledgements}
This research has been supported from the projects CEX2019-000918-M, CEX2020-001058-M (Unidades de Excelencia ``Mar\'{\i}a de Maeztu"), PID2019-110165GB-I00 and PID2020-118758GB-I00, financed by the spanish MCIN/ AEI/10.13039/501100011033/, as well as by the EU STRONG-2020 project, under the program  H2020-INFRAIA-2018-1 grant agreement no. 824093, and by PHAROS COST Action CA16214. H.K. acknowledges support from the PRE2020-093558 Doctoral Grant of the spanish MCIN/ AEI/10.13039/501100011033/.
L.T. also acknowledges support from the Generalitat Valenciana under contract PROMETEO/2020/023 and from the CRC-TR 211 'Strong-interaction matter under extreme conditions'- project Nr. 315477589 - TRR 211.

\section*{Data Availability}

The data underlying this article will be shared on reasonable request to the corresponding author.



\bibliographystyle{mnras}
\bibliography{reference} 

\begin{thebibliography}{}
\makeatletter
\relax
\def\mn@urlcharsother{\let\do\@makeother \do\$\do\&\do\#\do\^\do\_\do\%\do\~}
\def\mn@doi{\begingroup\mn@urlcharsother \@ifnextchar [ {\mn@doi@}
  {\mn@doi@[]}}
\def\mn@doi@[#1]#2{\def\@tempa{#1}\ifx\@tempa\@empty \href
  {http://dx.doi.org/#2} {doi:#2}\else \href {http://dx.doi.org/#2} {#1}\fi
  \endgroup}
\def\mn@eprint#1#2{\mn@eprint@#1:#2::\@nil}
\def\mn@eprint@arXiv#1{\href {http://arxiv.org/abs/#1} {{\tt arXiv:#1}}}
\def\mn@eprint@dblp#1{\href {http://dblp.uni-trier.de/rec/bibtex/#1.xml}
  {dblp:#1}}
\def\mn@eprint@#1:#2:#3:#4\@nil{\def\@tempa {#1}\def\@tempb {#2}\def\@tempc
  {#3}\ifx \@tempc \@empty \let \@tempc \@tempb \let \@tempb \@tempa \fi \ifx
  \@tempb \@empty \def\@tempb {arXiv}\fi \@ifundefined
  {mn@eprint@\@tempb}{\@tempb:\@tempc}{\expandafter \expandafter \csname
  mn@eprint@\@tempb\endcsname \expandafter{\@tempc}}}

\bibitem[\protect\citeauthoryear{Ahn et~al.}{Ahn et~al.}{2013}]{Ahn:2013poa}
Ahn J.~K.,  et~al., 2013, \mn@doi [Phys. Rev. C] {10.1103/PhysRevC.88.014003},
  88, 014003

\bibitem[\protect\citeauthoryear{Antoniadis et~al.}{Antoniadis
  et~al.}{2013}]{Antoniadis:2013pzd}
Antoniadis J.,  et~al., 2013, \mn@doi [Science] {10.1126/science.1233232}, 340,
  6131

\bibitem[\protect\citeauthoryear{Baiotti \& Rezzolla}{Baiotti \&
  Rezzolla}{2017}]{Baiotti:2016qnr}
Baiotti L.,  Rezzolla L.,  2017, \mn@doi [Rept. Prog. Phys.]
  {10.1088/1361-6633/aa67bb}, 80, 096901

\bibitem[\protect\citeauthoryear{Balberg \& Avraham}{Balberg \&
  Avraham}{1997}]{Balberg1997AnStrangeness}
Balberg S.,  Avraham G.,  1997, \mn@doi [Nuclear Physics A]
  {10.1016/S0375-9474(97)81465-0}, 625, 435

\bibitem[\protect\citeauthoryear{Baldo, Burgio  \& Schulze}{Baldo
  et~al.}{1998}]{Baldo:1998hd}
Baldo M.,  Burgio G.~F.,   Schulze H.~J.,  1998, \mn@doi [Phys. Rev. C]
  {10.1103/PhysRevC.58.3688}, 58, 3688

\bibitem[\protect\citeauthoryear{Bauswein, Janka  \& Oechslin}{Bauswein
  et~al.}{2010}]{Bauswein:2010dn}
Bauswein A.,  Janka H.~T.,   Oechslin R.,  2010, \mn@doi [Phys. Rev. D]
  {10.1103/PhysRevD.82.084043}, 82, 084043

\bibitem[\protect\citeauthoryear{Boguta \& Bodmer}{Boguta \&
  Bodmer}{1977}]{Boguta1977RelativisticSurface}
Boguta J.,  Bodmer A.~R.,  1977, \mn@doi [Nucl. Phys. A]
  {10.1016/0375-9474(77)90626-1}, 292, 413

\bibitem[\protect\citeauthoryear{Burgio, Schulze, Vidana  \& Wei}{Burgio
  et~al.}{2021}]{Burgio:2021vgk}
Burgio G.~F.,  Schulze H.~J.,  Vidana I.,   Wei J.~B.,  2021, \mn@doi [Prog.
  Part. Nucl. Phys.] {10.1016/j.ppnp.2021.103879}, 120, 103879

\bibitem[\protect\citeauthoryear{Burrows, Radice, Vartanyan, Nagakura, Skinner
  \& Dolence}{Burrows et~al.}{2020}]{Burrows:2019zce}
Burrows A.,  Radice D.,  Vartanyan D.,  Nagakura H.,  Skinner M.~A.,   Dolence
  J.,  2020, \mn@doi [Mon. Not. Roy. Astron. Soc.] {10.1093/mnras/stz3223},
  491, 2715

\bibitem[\protect\citeauthoryear{Camelio, Dietrich, Rosswog  \&
  Haskell}{Camelio et~al.}{2021}]{Camelio:2020mdi}
Camelio G.,  Dietrich T.,  Rosswog S.,   Haskell B.,  2021, \mn@doi [Phys. Rev.
  D] {10.1103/PhysRevD.103.063014}, 103, 063014

\bibitem[\protect\citeauthoryear{Chen \& Piekarewicz}{Chen \&
  Piekarewicz}{2014}]{Chen:2014sca}
Chen W.-C.,  Piekarewicz J.,  2014, \mn@doi [Phys. Rev. C]
  {10.1103/PhysRevC.90.044305}, 90, 044305

\bibitem[\protect\citeauthoryear{{Couch}}{{Couch}}{2017}]{2017RSPTA.37560271C}
{Couch} S.~M.,  2017, \mn@doi [Philos. Trans. Royal Soc. A]
  {10.1098/rsta.2016.0271}, 375, 20160271

\bibitem[\protect\citeauthoryear{Cromartie et~al.}{Cromartie
  et~al.}{2019}]{NANOGrav:2019jur}
Cromartie H.~T.,  et~al., 2019, \mn@doi [Nature Astron.]
  {10.1038/s41550-019-0880-2}, 4, 72

\bibitem[\protect\citeauthoryear{{Cusinato}, {Guercilena}, {Perego},
  {Logoteta}, {Radice}, {Bernuzzi}  \& {Ansoldi}}{{Cusinato}
  et~al.}{2022}]{2022EPJA...58...99C}
{Cusinato} M.,  {Guercilena} F.~M.,  {Perego} A.,  {Logoteta} D.,  {Radice} D.,
   {Bernuzzi} S.,   {Ansoldi} S.,  2022, \mn@doi [Eur. Phys. J A]
  {10.1140/epja/s10050-022-00743-5}, 58, 99

\bibitem[\protect\citeauthoryear{Demorest, Pennucci, Ransom, Roberts  \&
  Hessels}{Demorest et~al.}{2010}]{Demorest2010ShapiroStar}
Demorest P.,  Pennucci T.,  Ransom S.,  Roberts M.,   Hessels J.,  2010,
  \mn@doi [Nature] {10.1038/nature09466}, 467, 1081

\bibitem[\protect\citeauthoryear{Endrizzi, Logoteta, Giacomazzo, Bombaci,
  Kastaun  \& Ciolfi}{Endrizzi et~al.}{2018}]{Endrizzi:2018uwl}
Endrizzi A.,  Logoteta D.,  Giacomazzo B.,  Bombaci I.,  Kastaun W.,   Ciolfi
  R.,  2018, \mn@doi [Phys. Rev. D] {10.1103/PhysRevD.98.043015}, 98, 043015

\bibitem[\protect\citeauthoryear{Fischer, Whitehouse, Mezzacappa, Thielemann
  \& Liebendorfer}{Fischer et~al.}{2009}]{Fischer:2008rh}
Fischer T.,  Whitehouse S.~C.,  Mezzacappa A.,  Thielemann F.~K.,
  Liebendorfer M.,  2009, \mn@doi [Astron. Astrophys.]
  {10.1051/0004-6361/200811055}, 499, 1

\bibitem[\protect\citeauthoryear{Fonseca et~al.,}{Fonseca
  et~al.}{2016}]{Fonseca2016}
Fonseca E.,  et~al., 2016, \mn@doi [Astrophys. J.]
  {10.3847/0004-637X/832/2/167}, 832, 167

\bibitem[\protect\citeauthoryear{Friedman \& Gal}{Friedman \&
  Gal}{2021}]{Friedman:2021rhu}
Friedman E.,  Gal A.,  2021, \mn@doi [Phys. Lett. B]
  {10.1016/j.physletb.2021.136555}, 820, 136555

\bibitem[\protect\citeauthoryear{Glendenning}{Glendenning}{2000}]{Glendenning:2000}
Glendenning N.~K.,  2000, {Compact stars: Nuclear physics, particle physics,
  and general relativity}, 2 edn.
Springer, New York

\bibitem[\protect\citeauthoryear{Harada \& Hirabayashi}{Harada \&
  Hirabayashi}{2006}]{HARADA2006206}
Harada T.,  Hirabayashi Y.,  2006, \mn@doi [Nucl. Phys. A]
  {https://doi.org/10.1016/j.nuclphysa.2005.12.018}, 767, 206

\bibitem[\protect\citeauthoryear{Hotokezaka, Kiuchi, Kyutoku, Muranushi,
  Sekiguchi, Shibata  \& Taniguchi}{Hotokezaka
  et~al.}{2013}]{Hotokezaka:2013iia}
Hotokezaka K.,  Kiuchi K.,  Kyutoku K.,  Muranushi T.,  Sekiguchi Y.-i.,
  Shibata M.,   Taniguchi K.,  2013, \mn@doi [Phys. Rev. D]
  {10.1103/PhysRevD.88.044026}, 88, 044026

\bibitem[\protect\citeauthoryear{{Janka}}{{Janka}}{2017}]{2017hsn..book.1575J}
{Janka} H.-T.,  2017, in {Alsabti} A.~W.,  {Murdin} P.,  eds, , Handbook of
  Supernovae.
p.~1575, \mn@doi{10.1007/978-3-319-21846-5\_4}

\bibitem[\protect\citeauthoryear{Janka, Langanke, Marek, Martinez-Pinedo  \&
  Mueller}{Janka et~al.}{2007}]{Janka:2006fh}
Janka H.-T.,  Langanke K.,  Marek A.,  Martinez-Pinedo G.,   Mueller B.,  2007,
  \mn@doi [Phys. Rept.] {10.1016/j.physrep.2007.02.002}, 442, 38

\bibitem[\protect\citeauthoryear{Kohno, Fujiwara, Watanabe, Ogata  \&
  Kawai}{Kohno et~al.}{2006}]{Kohno:2006iq}
Kohno M.,  Fujiwara Y.,  Watanabe Y.,  Ogata K.,   Kawai M.,  2006, \mn@doi
  [Phys. Rev. C] {10.1103/PhysRevC.74.064613}, 74, 064613

\bibitem[\protect\citeauthoryear{Lattimer \& Swesty}{Lattimer \&
  Swesty}{1991}]{Lattimer:1991nc}
Lattimer J.~M.,  Swesty F.~D.,  1991, \mn@doi [Nucl. Phys. A]
  {10.1016/0375-9474(91)90452-C}, 535, 331

\bibitem[\protect\citeauthoryear{Mezzacappa et~al.}{Mezzacappa
  et~al.}{2015}]{mezzapaca:corrected}
Mezzacappa A.,  et~al., 2015, Numerical Modeling of Space Plasma Flows:
  ASTRONUM-2014 ASP Conference Series, 498, 108

\bibitem[\protect\citeauthoryear{Miller et~al.}{Miller
  et~al.}{2019}]{Miller:2019cac}
Miller M.~C.,  et~al., 2019, \mn@doi [Astrophys. J. Lett.]
  {10.3847/2041-8213/ab50c5}, 887, L24

\bibitem[\protect\citeauthoryear{Miller et~al.}{Miller
  et~al.}{2021}]{Miller:2021qha}
Miller M.~C.,  et~al., 2021, \mn@doi [Astrophys. J. Lett.]
  {10.3847/2041-8213/ac089b}, 918, L28

\bibitem[\protect\citeauthoryear{Noumi et~al.}{Noumi
  et~al.}{2002}]{Noumi:2001tx}
Noumi H.,  et~al., 2002, \mn@doi [Phys. Rev. Lett.]
  {10.1103/PhysRevLett.89.072301}, 89, 072301

\bibitem[\protect\citeauthoryear{Oertel, Hempel, Kl\"ahn  \& Typel}{Oertel
  et~al.}{2017}]{Oertel:2016bki}
Oertel M.,  Hempel M.,  Kl\"ahn T.,   Typel S.,  2017, \mn@doi [Rev. Mod.
  Phys.] {10.1103/RevModPhys.89.015007}, 89, 015007

\bibitem[\protect\citeauthoryear{O’Connor \& Couch}{O’Connor \&
  Couch}{2018}]{OConnor2018ExploringSupernovae}
O’Connor E.~P.,  Couch S.~M.,  2018, \mn@doi [Astrophys. J.]
  {10.3847/1538-4357/AADCF7}, 865, 81

\bibitem[\protect\citeauthoryear{Pandharipande}{Pandharipande}{1971}]{Pandharipande:1971up}
Pandharipande V.~R.,  1971, \mn@doi [Nucl. Phys. A]
  {10.1016/0375-9474(71)90193-X}, 178, 123

\bibitem[\protect\citeauthoryear{Pascal, Novak  \& Oertel}{Pascal
  et~al.}{2022}]{Pascal:2022qeg}
Pascal A.,  Novak J.,   Oertel M.,  2022, \mn@doi [Mon. Not. Roy. Astron. Soc.]
  {10.1093/mnras/stac016}, 511, 356

\bibitem[\protect\citeauthoryear{Pons, Reddy, Prakash, Lattimer  \&
  Miralles}{Pons et~al.}{1999}]{Pons1999EvolutionStars}
Pons J.~A.,  Reddy S.,  Prakash M.,  Lattimer J.~M.,   Miralles J.~A.,  1999,
  \mn@doi [Astrophys. J.] {10.1086/306889}, 513, 780

\bibitem[\protect\citeauthoryear{Raduta}{Raduta}{2022}]{Raduta:2022elz}
Raduta A.~R.,  2022.  (\mn@eprint {arXiv} {2205.03177})

\bibitem[\protect\citeauthoryear{Raduta, Oertel  \& Sedrakian}{Raduta
  et~al.}{2020}]{Raduta:2020fdn}
Raduta A.~R.,  Oertel M.,   Sedrakian A.,  2020, \mn@doi [Mon. Not. Roy.
  Astron. Soc.] {10.1093/mnras/staa2491}, 499, 914

\bibitem[\protect\citeauthoryear{Raduta, Nacu  \& Oertel}{Raduta
  et~al.}{2021}]{Raduta:2021coc}
Raduta A.~R.,  Nacu F.,   Oertel M.,  2021, \mn@doi [Eur. Phys. J. A]
  {10.1140/epja/s10050-021-00628-z}, 57, 329

\bibitem[\protect\citeauthoryear{Raithel, Paschalidis  \& {\"{O}}zel}{Raithel
  et~al.}{2021}]{Raithel2021RealisticSimulations}
Raithel C.,  Paschalidis V.,   {\"{O}}zel F.,  2021, \mn@doi [Phys. Rev. D]
  {10.1103/PhysRevD.104.063016}, 104, 063016

\bibitem[\protect\citeauthoryear{Riley et~al.}{Riley
  et~al.}{2019}]{Riley:2019yda}
Riley T.~E.,  et~al., 2019, \mn@doi [Astrophys. J. Lett.]
  {10.3847/2041-8213/ab481c}, 887, L21

\bibitem[\protect\citeauthoryear{Riley et~al.}{Riley
  et~al.}{2021}]{Riley:2021pdl}
Riley T.~E.,  et~al., 2021, \mn@doi [Astrophys. J. Lett.]
  {10.3847/2041-8213/ac0a81}, 918, L27

\bibitem[\protect\citeauthoryear{Rosswog}{Rosswog}{2015}]{Rosswog:2015nja}
Rosswog S.,  2015, \mn@doi [Int. J. Mod. Phys. D] {10.1142/S0218271815300128},
  24, 1530012

\bibitem[\protect\citeauthoryear{Saha et~al.}{Saha et~al.}{2004}]{Saha:2004ha}
Saha P.~K.,  et~al., 2004, \mn@doi [Phys. Rev. C] {10.1103/PhysRevC.70.044613},
  70, 044613

\bibitem[\protect\citeauthoryear{Sedrakian \& Harutyunyan}{Sedrakian \&
  Harutyunyan}{2021}]{Sedrakian:2021qjw}
Sedrakian A.,  Harutyunyan A.,  2021, \mn@doi [Universe]
  {10.3390/universe7100382}, 7, 382

\bibitem[\protect\citeauthoryear{Serot \& Walecka}{Serot \&
  Walecka}{1986}]{Serot:1984ey}
Serot B.~D.,  Walecka J.~D.,  1986, Adv. Nucl. Phys., 16, 1

\bibitem[\protect\citeauthoryear{Serot \& Walecka}{Serot \&
  Walecka}{1997}]{Serot1997RecentHadrodynamics}
Serot B.~D.,  Walecka J.~D.,  1997, \mn@doi [Int. J. Mod. Phys. E]
  {10.1142/S0218301397000299}, 6, 515

\bibitem[\protect\citeauthoryear{Shen, Toki, Oyamatsu  \& Sumiyoshi}{Shen
  et~al.}{1998a}]{Shen:1998by}
Shen H.,  Toki H.,  Oyamatsu K.,   Sumiyoshi K.,  1998a, \mn@doi [Prog. Theor.
  Phys.] {10.1143/PTP.100.1013}, 100, 1013

\bibitem[\protect\citeauthoryear{Shen, Toki, Oyamatsu  \& Sumiyoshi}{Shen
  et~al.}{1998b}]{Shen:1998gq}
Shen H.,  Toki H.,  Oyamatsu K.,   Sumiyoshi K.,  1998b, \mn@doi [Nucl. Phys.
  A] {10.1016/S0375-9474(98)00236-X}, 637, 435

\bibitem[\protect\citeauthoryear{Sumiyoshi, Yamada  \& Suzuki}{Sumiyoshi
  et~al.}{2007}]{Sumiyoshi2007DynamicsDependence}
Sumiyoshi K.,  Yamada S.,   Suzuki H.,  2007, \mn@doi [Astrophys. J.]
  {10.1086/520876}, 667, 382

\bibitem[\protect\citeauthoryear{Takahashi et~al.}{Takahashi
  et~al.}{2001}]{Takahashi:2001nm}
Takahashi H.,  et~al., 2001, \mn@doi [Phys. Rev. Lett.]
  {10.1103/PhysRevLett.87.212502}, 87, 212502

\bibitem[\protect\citeauthoryear{Tolos, Centelles  \& Ramos}{Tolos
  et~al.}{2017a}]{Tolos:2017lgv}
Tolos L.,  Centelles M.,   Ramos A.,  2017a, \mn@doi [Publ. Astron. Soc.
  Austral.] {10.1017/pasa.2017.60}, 34, e065

\bibitem[\protect\citeauthoryear{Tolos, Centelles  \& Ramos}{Tolos
  et~al.}{2017b}]{Tolos:2016hhl}
Tolos L.,  Centelles M.,   Ramos A.,  2017b, \mn@doi [Astrophys. J.]
  {10.3847/1538-4357/834/1/3}, 834, 3

\bibitem[\protect\citeauthoryear{Typel et~al.}{Typel
  et~al.}{2022}]{Typel:2022lcx}
Typel S.,  et~al., 2022.  (\mn@eprint {arXiv} {2203.03209})

\bibitem[\protect\citeauthoryear{Vida{\~{n}}a, Polls, Ramos, Engvik  \&
  Hjorth-Jensen}{Vida{\~{n}}a et~al.}{2000}]{Vidana2000Hyperon-hyperonMatter}
Vida{\~{n}}a I.,  Polls A.,  Ramos A.,  Engvik L.,   Hjorth-Jensen M.,  2000,
  \mn@doi [Phys. Rev. C] {10.1103/PhysRevC.62.035801}, 62, 8

\bibitem[\protect\citeauthoryear{Vidana, Bombaci, Polls  \& Ramos}{Vidana
  et~al.}{2003}]{Vidana:2002rg}
Vidana I.,  Bombaci I.,  Polls A.,   Ramos A.,  2003, \mn@doi [Astron.
  Astrophys.] {10.1051/0004-6361:20021840}, 399, 687

\bibitem[\protect\citeauthoryear{Walecka}{Walecka}{1974}]{Walecka:1974qa}
Walecka J.~D.,  1974, \mn@doi [Annals Phys.] {10.1016/0003-4916(74)90208-5},
  83, 491

\makeatother
\end{thebibliography}








\bsp	
\label{lastpage}
\end{document}